# Temperature dependence of Fano resonance in nanodiamonds synthesized at high static pressures

A.A.Shiryaev[+*], E.A.Ekimov[×], V.Yu. Prokof'ev[*], M.V.Kondrin[×1]

[+]Frumkin Institute of Physical Chemistry and Electrochemistry, Russian Academy of Sciences, Moscow, 119071 Russia

[*]Institute of Geology of Ore Deposits, Petrography, Mineralogy, and Geochemistry (IGEM), Russian Academy of Sciences, Moscow, 119017 Russia

[×]Vereshchagin Institute for High Pressure Physics, Russian Academy of Sciences, Moscow 108840, Russia

Corresponding author e-mail: mkondrin@hppi.troitsk.ru

**Abstract.** Temperature dependence of Fano resonance, recently discovered in infra-red (IR) spectra of nanodiamonds synthesized from chloroadamantane at high static pressures, is investigated. For the first time, marked variations of the resonance parameters are observed. On heating, the shape of the Fano resonance changes considerably; the effect completely disappears above 350 °C, but is recovered after cooling to ambient conditions. Such behavior implies that assignment of the Fano effect to the surface transfer doping mechanism is not very plausible for the studied samples. The resonance shape varies due to strong temperature dependence of the difference of frequencies of IR-active "bright" and Raman-active "dark" modes of nanodiamond. The frequency of the Raman "dark" mode is only weakly temperature-dependent.

**1. Introduction**. The Fano effect predicted by Ugo Fano in 1930-ies [1,2] attracts considerable attention in recent years [3-5]. Rapid switching from absorption to transmission induced by the Fano effect is of significant basic and applied interest. Fano resonances are observed in metamaterials, metasurfaces, nanoshells and many other types of nanodisperse materials [3-9].

Recently, two groups reported observation of Fano effect in the infra-red (IR) absorption spectra of nanodiamonds with sizes larger than 2 nm, synthesized from adamantane and its derivatives at high static pressures [10, 11]. Note, that synthesis from halogenated hydrocarbons is a rapidly developing technology [12-16], allowing growth of nanodiamond grains with well-controlled sizes from 1 nm and larger. Such nanodiamonds are promising for applications in biomedicine, quantum optics and cryptography [17]. In case of nanodiamonds, Fano resonance is manifested as a "transparency" window at approx. 1330 cm$^{-1}$, i.e. in the vicinity of diamond Raman frequency. Recall that in pure diamond the Raman mode is IR-inactive. The origin of Fano resonance in nanodiamond remains debatable. The IR absorption at the Raman frequency is tentatively linked to conductive states on surfaces of nanodiamond grains. The appearance of these states is not yet fully understood: they might be related to adsorbed water/oxygen/etc. (so-called surface transfer doping mechanism [18-20]) on hydrogenated nanodiamonds [10] or arise from surface reconstruction [11]. In the latter case, the reconstruction leads to formation of trans-polyacethylene-like fragments on the surfaces. A correlation between the electrical conductivity of nanodiamonds and appearance of the Fano effect in IR absorption was established earlier: the 8 nm grains display Fano effect and significant conductivity ($10^5$-$10^6$ Ohm·cm), whereas for the grains smaller than 2 nm both these features are absent [11]. To elucidate origin of conductive states in nanodiamonds demonstrating Fano effect at room temperature, we have performed in situ measurements of IR reflection spectra in a broad temperature range both in inert atmosphere (dry $N_2$ flow) and in air. These results are presented below.

Fano effect consists of coherent interaction of an optically-active ("bright") broad-band mode with a narrow optically inactive ("dark") mode. Due to this interaction the "dark" mode is manifested in optical spectra as a peak with a peculiar asymmetric lineshape. Despite quantum origin of Fano resonance, the resulting effects can be modelled using two classic coupled oscillators, which can be described using matrix equation [5, 9, 21, 22]:

$$\begin{pmatrix} \omega - \omega_1 - i\gamma & g \\ g & \omega - \omega_2 \end{pmatrix} \cdot \begin{pmatrix} x_1 \\ x_2 \end{pmatrix} = i \begin{pmatrix} f_1 \\ 0 \end{pmatrix},$$

where $\omega_1$ and $\omega_2$ – are characteristic frequencies of the "bright" and "dark" modes, respectively, $\gamma$ – width of the "bright" mode, $g$ – coupling constant between the "bright" and "dark" modes. Dimension of these constants, of driving force amplitude $f_1$ and frequency $\omega$ are given in energy units. The system response $x_i$ is dimensionless. Note that the driving force consists of a single component $f_1$, implying that only one of the oscillators directly interacts with the external field.

The solution of the above equation is expressed by the function (1):

$$\| x_1 \|^2 = \frac{\| f_1 \|^2 A(\omega - \omega_0 + \Gamma F)^2}{(\omega - \omega_0)^2 + \Gamma^2}, \tag{1}$$

where $A$, $\Gamma$, $F$, $\omega_0$ – functions, depending on frequency $\omega$:

$$F = \frac{\omega - \omega_1}{\gamma}, \tag{2}$$

$$\Gamma = \frac{g^2}{\gamma(1 + F^2)}, \tag{3}$$

$$\omega_0 = \omega_2 + \Gamma F, \tag{4}$$

$$A = \frac{1}{\gamma^2(1 + F^2)}. \tag{5}$$

At the same time, in the vicinity of the "dark" mode frequency (infinitely narrow in the coupled oscillators model) substitution $\omega \to \omega_2$ in the equation (2) allows approximation of these functions by constants. In this case the equation (1) becomes asymmetric, thus giving the characteristic Fano line shape. Accrodingly, the parameter $A$ is the absorption amplitude, $\Gamma$ and $\omega_0$ – width and characteristic frequency of the absorption line and the Fano parameter, $F$, is the line asymmetry coefficient. Typical example of such line shape is shown in Fig. 1, where the experimental IR absorption spectrum of 8 nm nanodiamonds synthesised from chloroadamantane at high pressures is fitted. Note that the line shape asymmetry leads to appearance of the "transmission window" at frequencies above the Fano resonance.

An important note about the fitting shown in Fig. 1 should be made. We have employed the equation (1) complemented by an additional incoherent background approximated by a straight line. The equation (1) implies positive value of the parameter $A$ for the absorption line. In course of unrestricted fit of experimental data this proportionality coefficient may turn negative. In such cases an additional spectral density can be added following the formula:

$$-\frac{(F+\Omega)^2}{(1+\Omega^2)} + (1+F^2) = F^2\frac{(1/F-\Omega)^2}{(1+\Omega^2)}, \tag{6}$$

where $\Omega = (\omega - \omega_0)/\Gamma$. Consequently, the amplitude of the Fano resonance (equal to -1) can be converted into a positive number ($F^2$) with simultaneous substitution $F \to -1/F$. We note that in our previous article [11] this procedure was not applied (the resonance amplitude was negative), leading to wrong value of the Fano parameter. The Figure 1 shows corrected values of the $F$ parameter.

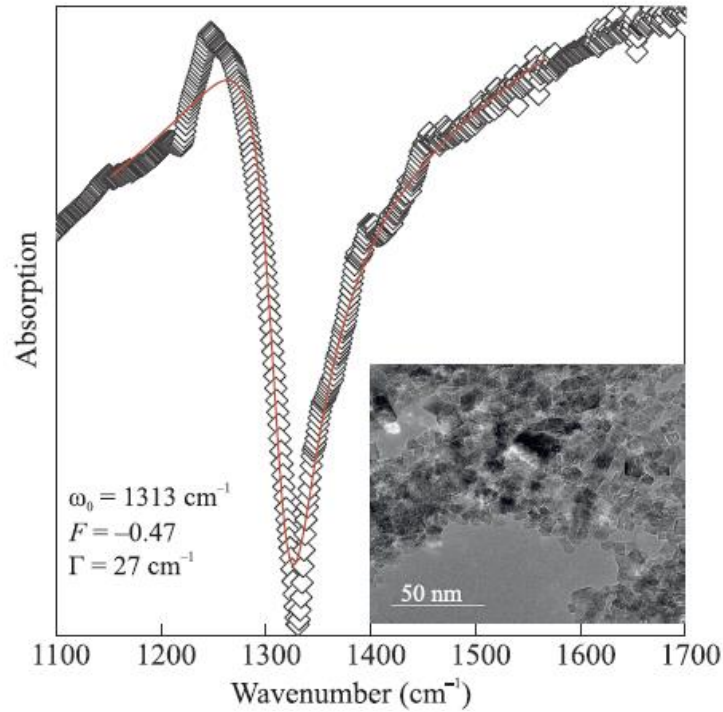

**Figure 1.** Fit (red line) of experimental IR spectrum of 8 nm nanodiamond sample (empty diamonds) using Equation (1) and a linear incoherent background. Obtained values of non-linear parameters are indicated. Inset: transmission electron microscopy image of the studied sample.

**2. Experimental.** Spectroscopic investigations of nanodiamonds with 8 nm grain size (see also [11]) were performed at the Geochemistry lab of IGEM RAS. Nanodiamond powder was measured on an Al mirror placed into Linkam THMSG600 heating stage [23] powered by a T95 temperature controller [24]. The mirror with the sample was covered by a IR-transparent

BaF$_2$ plate having a direct contact with the heater, thus assuring uniform equilibrium heating. The temperature was calibrated using a reference phase transition at 289 °C; the measured temperature was 288 °C. Therefore, the accuracy of the temperature readings was ±1 °C, which is comparable with earlier studies of mineralogical thin sections [25]. IR spectra were recorded in reflection geometry using Lumos II FTIR microscope at a spectral resolution of 2 cm$^{-1}$ with 600-1000 scans (5–10 min). The measurements were performed in air between 22 and 400 °C, and in inert atmosphere of continuous flow of dry N$_2$ between -80 and 380 °C with temperature steps 20 - 50 °C. The spectra were also recorded at room temperature after the heating cycle. Comparison of several independent runs showed that the results were insensitive to the heating regime.

The absence of the Fano effect for nanodiamonds smaller than 2 nm was shown in [11]. In these samples in vicinity of the Raman frequency several IR absorption peaks are observed. In the intermediate size range the Fano resonance is heavily modulated by IR absorption bands [11]. Nanodiamonds with grain size 8 nm demonstrate well-pronounced Fano effect; this sample was used in the current work for the step heating spectroscopic experiments.

**3. Results.** Experimental IR reflection spectra recorded at temperatures between -80 – -400 °C are shown in Figs. 2, 3. At high temperatures the Fano resonance disappears. Only the Fano band at ~1330 cm$^{-1}$ and valence vibrations of C-H bonds at ≈3000 cm$^{-1}$ are observed in spectra, no other bands are present. The C-H groups are common for nanodiamonds. Irreversible surface reconstruction and disappearence of surface charge contribution to electrical conductivity are observed for hydrogenated diamond films after heating in air at 230 °C [18]. This effect is obviously related to removal of physisorbed water/oxygen molecules, since such low temperatures do not affect H-containing functional groups [26]. In our experiments, cooling from 380-400 °C to room temperature leads to recovery of the Fano effect both in air and in inert atmosphere. Persistence of the Fano effect in nanodiamonds after prolonged (24 h) heating at 250 °C in air was demonstrated earlier [11].

Prior to discussion of the Fano effect on heated and annealed nanodiamond samples, lets discuss amplitude of the bands in the reflection spectra. One might assume that the absorption coefficient does not change with temperature with accuracy down to a multiplicative constant.

However, in our experiments the amplitude of the C-H band increases with temperature (see also [27]). Such behaviour may be explained by changes in sample emissivity and/or by alterations of the specimen surface (e.g., redistibution of nanograins). Consequently, in our work the experimental data shown in Figs. 2,3 are normalised to the C-H band amplitude. In any case, the variations in the multiplicative constant in the studied temperature range are moderate, do not exceed 2-3 times and do not influence subsequent evaluation, accurate to an order of magnitude.

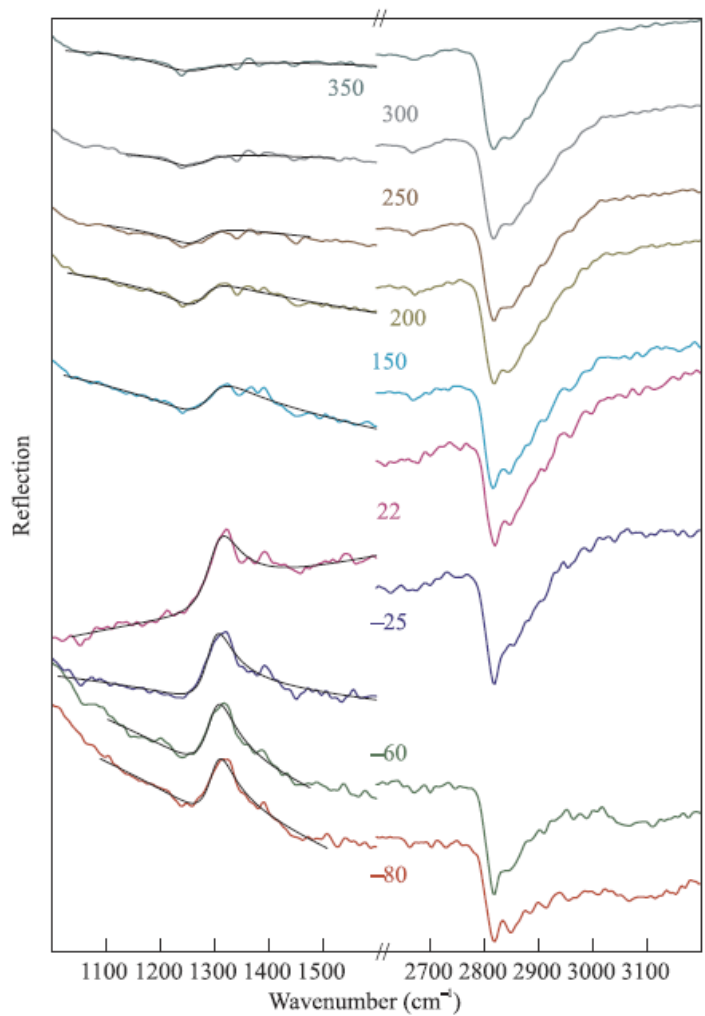

**Figure 2.** IR reflection spectra of 8 nm nanodiamond sample recorded in situ during cooling/heating in $N_2$ medium; temperatures are in °C. Fits of Fano resonance using Equation (1) with linear incoherent background are shown by solid black lines.

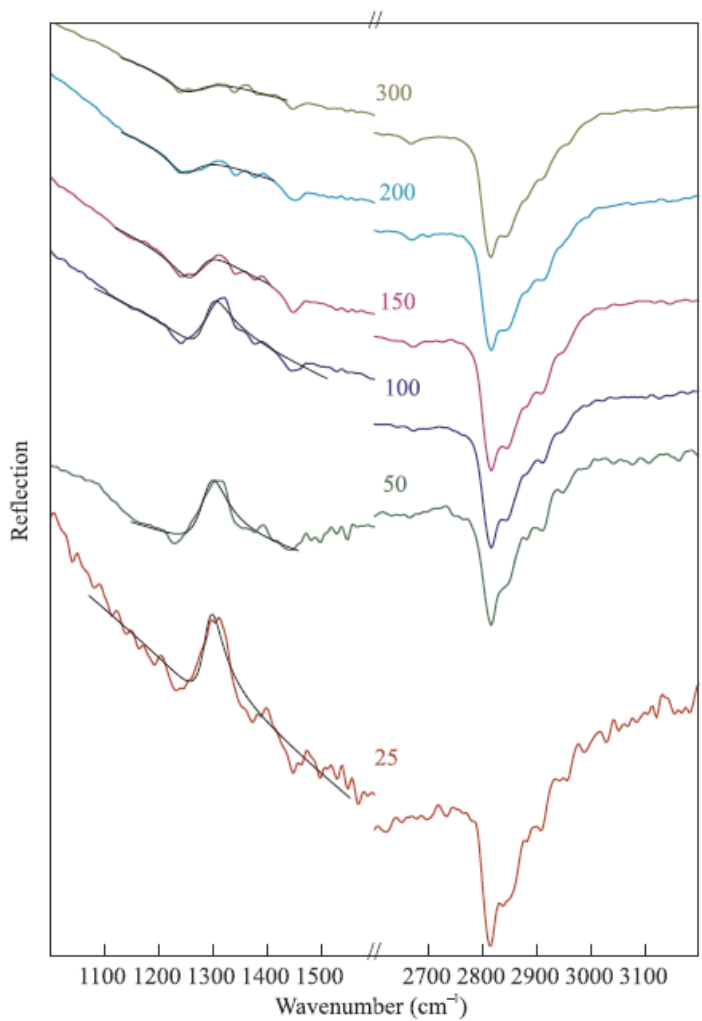

**Figure 3.** IR reflection spectra of 8 nm nanodiamond sample recorded in situ during heating in air; temperatures are in °C. Fits of Fano resonance using Equation (1) with linear incoherent background are shown by solid black lines.

Solid black lines in the left part of Figs. 2,3 show results of fitting of the "transmission window" using Equation (1). Despite marked changes in the line position and shape in the temperature range -80 – -300 °C, it is possible to obtain smooth variations of the fit parameters; some of the obtained vaules are shown in Table 1 and Fig. 4. It is still possible to make the fit of the spectrum at 350 °C, but the parameters became unphysical. At yet higher temperatures the fit procedure fails. In our view this behaviour indicates (reversible) destruction of the Fano resonance above 350 °C.

**Table 1.** Parameters of Fano resonance for experimental data at different temperatures ($T$). $A$ – scattering amplitude, $\Gamma$ and $\omega_0$ – width and characteristic frequency of the absorption line, $F$ – Fano parameter, $A_{corr}$ – band amplitude corrected for the C-H bands amplitude (see text for detail). The upper part of the table corresponds to experiment in $N_2$ medium, the lower – in air.

| $T$, °C | $\omega_0$, cm$^{-1}$ | $F$ | $\Gamma$, cm$^{-1}$ | $A$ | $A_{corr}$ |
|---|---|---|---|---|---|
| −80 | 1301.1 | −0.457 | 28.4 | 0.006085 | 0.50716 |
| −60 | 1298.9 | −0.418 | 31.4 | 0.004372 | 0.54660 |
| −25 | 1296.6 | −0.419 | 29.4 | 0.004077 | 0.48777 |
| 22 | 1303.8 | −0.344 | 36.6 | 0.010737 | 1.65199 |
| 150 | 1291.6 | −0.911 | 46.4 | 0.002407 | 0.17193 |
| 200 | 1275.6 | −1.545 | 38.4 | 0.001177 | 0.07362 |
| 250 | 1268.8 | −2.310 | 47.0 | 0.000271 | 0.01170 |
| 300 | 1260.0 | −4.666 | 45.4 | 0.000145 | 0.00608 |
| | | | | | |
| 25 | 1292.1 | −0.393 | 21.7 | 0.005289 | 0.66112 |
| 50 | 1291.8 | −0.355 | 29.6 | 0.003902 | 0.48778 |
| 100 | 1290.9 | −0.658 | 27.3 | 0.005710 | 0.30054 |
| 150 | 1270.4 | −1.767 | 32.2 | 0.001379 | 0.06571 |
| 200 | 1250.5 | −7.124 | 36.4 | 0.000108 | 0.00387 |
| 300 | 1255.7 | −5.982 | 46.3 | 0.000142 | 0.00475 |

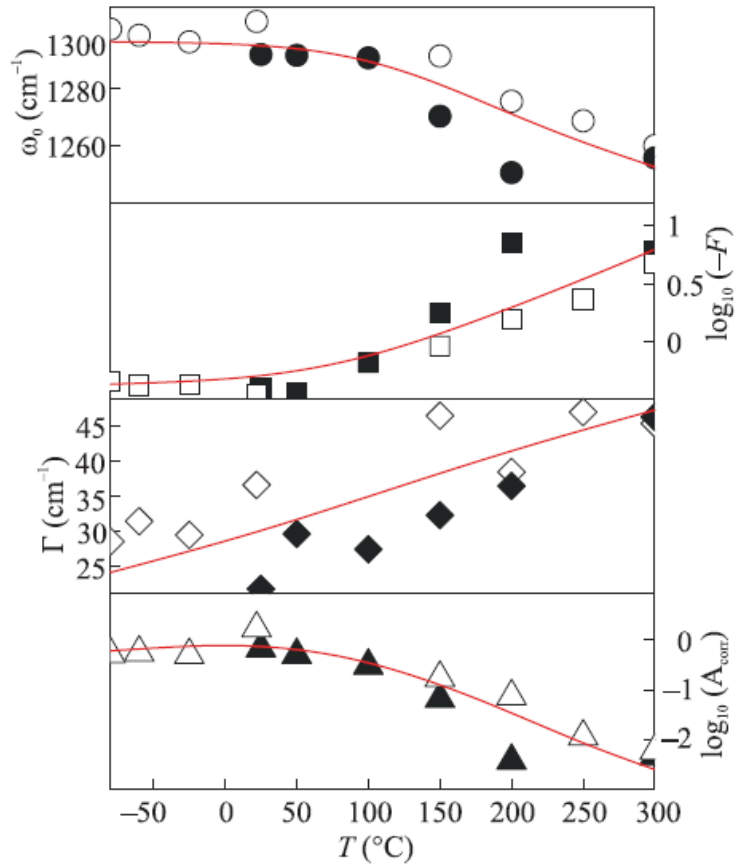

**Figure 4.** Temperature dependence of parameters from experimentally observed Fano resonance (see Figs. 2,3) using fit with Equation 1. Solid red lines indicate general trends. Black symbols – experiment in air; empty diamonds – in $N_2$ medium.

Examination of the fit results shows that the Fano parameter $F$ and $A$, $A_{corr}$ values change considerably, for 1 and 1.5-2 orders of magnitude, respectively. Changes in the characteristic frequency $\omega_0$ and linewidth $\Gamma$ are relatively modest. Lets consider correspondence of these variations of the experimental parameters with properties of bound oscillators model. To the best of our knowledge, despite intense discussions in literature, this model was not yet applied for qualitative evaluation of experimental data; misprints in published equations are partly responsible for this situation.

Examination of equations (2-5) shows that non-linear fit parameters ($\omega_0$, $F$, $\Gamma$), expressed in absolute units, allow to calculate only the value of characteristic frequency of the

"dark" mode $\omega_2$ (upper graph of Fig. 5). Figure 5 shows that in the temperature range -80 – -200 °C this frequency coincides with diamond Raman mode ($\pm 10$ cm$^{-1}$), implying that the Raman mode is involved in the Fano effect in nanodiamonds. Determination of other parameteres of the model implies evaluation of the temperature dependence of the $\gamma$ (width of the "bright" mode) from the dependence of the Fano resonance amplitude $A$ (Eq. (5)). As noted above, the latter is known from experimental data (within a multiplication factor); in our work we use the amplitude $A_{corr}$ normalised to the C-H vibrations amplitude. This approach allows to determine the strongest relative change of any parameter of the model. These values are shown in Fig. 5 in arbitrary units.

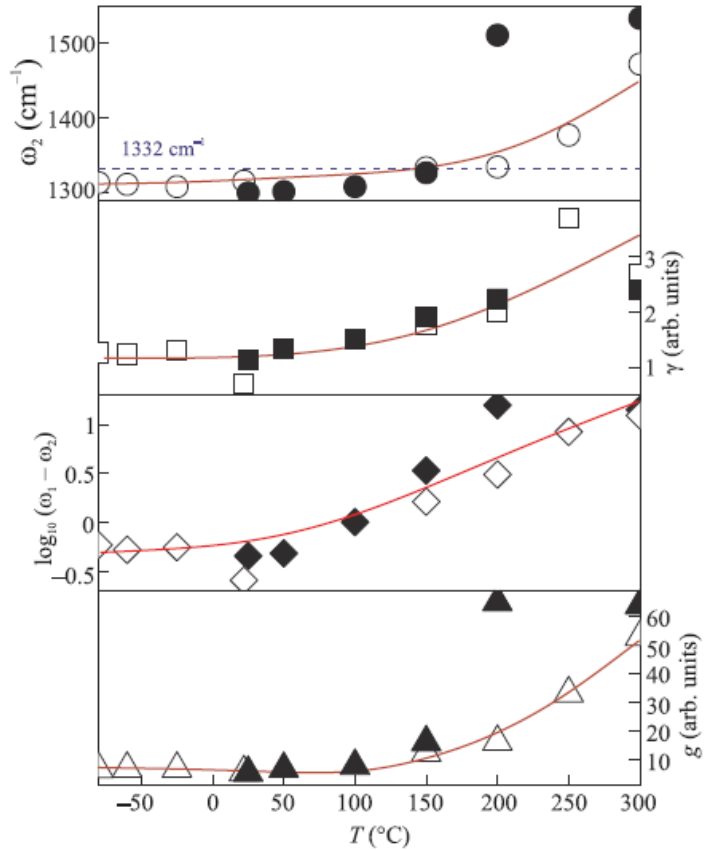

**Figure 5.** Temperature dependence of parameters of two coupled oscillators model. $\omega_2$ – absolute value of the "dark" mode frequency (dashed line shows position of diamond Raman line); $\omega_1 - \omega_2$ detuning of the "dark" and "bright" modes; $\gamma$ – width of the "bright" mode; $g$ – coupling constant of the "dark" and "bright" modes interaction. Solid red lines indicate general trends. Black symbols – experiment in air; empty diamonds – in N$_2$ medium.

Figure 5 shows that the strongest temperature effect – almost 25 times – is observed for the difference between frequencies of the "bright" and "dark" modes $\omega_2 - \omega_1$ (detuning). Less pronounced change (~8 times) occurs for the coupling constant $g$ between these modes. Therefore, one concludes that the main factor responsible for the temperature-induced change in the the Fano resonance line shape is the shift of the "bright" mode frequency. In experimental spectra this band is broad, precluding precise determination of its position. At the same, the changes in $\omega_2 - \omega_1$ influence the coupling constant $g$ of the "bright" and "dark" modes. Presumably, the position of the "bright" mode is close to that of the "dark" one, which, as shown above, is in the vicinity of the diamond Raman mode. This suggestion stems from small magnitude of the experimentally observed Fano parameter. We note also that for larger nanodiamonds (~30 nm) the detuning factor $\omega_2 - \omega_1$ nearly vanishes [10], giving the $F$ value close to 0 and almost symmetric transmission window.

**4. Conclusions.** Temperature dependence of Fano effect in nanodiamonds synthesised from chloroadamantane at high static pressures is investigated. Significant temperature-induced changes in Fano resonance in nanodiamonds – the Fano parameter varies for almost an order of magnitude – are observed for the first time. Heating of the nanodiamonds to 350 °C both in air and in inert atmosphere ($N_2$ flow) destroys the Fano effect; however, it is recovered upon cooling to room temperature. Such behaviour suggests the surface transfer doping mechanism is an unlikely reason for the appearance of the Fano resonance in the studied samples. Experimental data were used to investigate temperature behaviour of various parameters from the Fano effect equation. It is shown that the most important variations are observed for the detuning between the "dark" and "bright" IR-active modes, which changes for more than order of magnitude in the studied temperature range. The frequency of the "dark" mode at temperatures up to 200 °C coincides with diamond Raman mode with a good accuracy.

We acknoledge support from Russian Foundation for Basic Research (grant #20-52-26017) and Dr. Stepan Stehlik for discussions.

## References

[1] U. Fano, *Il Nuovo Cimento* (1924-1942) **12**, 154 (1935).

[2] U. Fano, *Phys. Rev.* **124**, 1866 (1961).

[3] B. Luk'yanchuk, N. I. Zheludev, S. A. Maier, N. J. Halas, P. Nordlander, H. Giessen, and C. T. Chong, *Nature Mater.* **9**, 707 (2010).

[4] F. J. Garca de Abajo, *Rev. Mod. Phys.* **79**, 1267 (2007).

[5] M. F. Limonov, M. V. Rybin, A. N. Poddubny, and Y. S. Kivshar, *Nature Photon.* **11**, 543 (2017).

[6] M. I. Tribelsky and A. E. Miroshnichenko, *Phys.-Uspekhi* **65**, 40 (2022).

[7] F. Lapointe, E. Gaufrès, I. Tremblay, N. Y.-W. Tang, R. Martel, and P. Desjardins, *Phys. Rev. Lett.* **109**, 097402 (2012).

[8] P. Gu, X. Cai, G. Wu, C. Xue, J. Chen, Z. Zhang, Z. Yan, F. Liu, C. Tang, W. Du, Z. Huang, and Z. Chen, *Nanomaterials* **11**, 2039 (2021).

[9] M. F. Limonov, *Adv. Opt. Photonics* **13**, 703 (2021).

[10] O. S. Kudryavtsev, R. H. Bagramov, A. M. Satanin, A. A. Shiryaev, O. I. Lebedev, A. M. Romshin, D. G. Pasternak, A. V. Nikolaev, V. P. Filonenko, and I. I. Vlasov, *Nano Lett.* **22**, 2589 (2022).

[11] E. Ekimov, A. A. Shiryaev, Y. Grigoriev, A. Averin, E. Shagieva, S. Stehlik, and M. Kondrin, *Nanomaterials* **12**, 351 (2022).

[12] V. A. Davydov, A. V. Rakhmanina, S. G. Lyapin, I. D. Ilichev, K. N. Boldyrev, A. A. Shiryaev, and V. N. Agafonov, *JETP Lett.* **99**, 585 (2014).

[13] E. Ekimov, S. Lyapin, Y. Grigoriev, I. Zibrov, and K. Kondrina, *Carbon* **150**, 436 (2019).

[14] E. Ekimov, M. Kondrin, S. Lyapin, Y. Grigoriev, A. Razgulov, V. Krivobok, S. Gierlotka, and S. Stelmakh, *Diam. Relat. Mater.* **103**, 107718 (2020).

[15] E. Ekimov, K. Kondrina, I. Zibrov, S. Lyapin, M. Lovygin, and P. Kazanskiy, *Materials Research Bulletin* **137**, 111189 (2021).

[16] M. V. Kondrin, I. P. Zibrov, S. G. Lyapin, Y. V. Grigoriev, R. A. Khmelnitskiy, and E. A. Ekimov, *ChemNanoMat* **7**, 17 (2021).

[17] E. A. Ekimov and M. V. Kondrin, *Phys.-Uspekhi* **60**, 539 (2017).

[18] F. Maier, M. Riedel, B. Mantel, J. Ristein, and L. Ley, *Phys. Rev. Lett.* **85**, 3472 (2000).

[19] W. Chen, D. Qi, X. Gao, and A. T. S. Wee, *Progress in Surface Science* **84**, 279 (2009).

[20] K. G. Crawford, I. Maini, D. A. Macdonald, and D. A. Moran, *Progress in Surface Science* **96**, 100613 (2021).

[21] Y. S. Joe, A. M. Satanin, and C. S. Kim, *Phys. Scr.* **74**, 259 (2006).

[22] B. Gallinet and O. J. F. Martin, *Phys. Rev. B* **83**, 235427 (2011).

[23] THMSG600 Temperature Controlled Geology Stage. User Guide, Linkam Scientific Instruments, https://linkamscientific.squarespace.com/ archivemanuals.

[24] PE95/T95 System Controller. User Guide, Linkam Scientific Instruments, https://linkamscientific.squarespace.com/ archivemanuals.

[25] V. Y. Prokof'ev, I. A. Baksheev, F. Y. Korytov, and J. Touret, *Comptes Rendus Geoscience* **338**, 617 (2006).

[26] A. P. Koscheev, Gas desorption from detonation nanodiamonds during temperature-programmed pyrolysis, in *Carbon Nanomaterials for Gas Adsorpton*, Pan Stanford Publishing Pte. Ltd., N.Y. (2012), p. 219.

[27] A. Maturilli, A. A. Shiryaev, I. I. Kulakova, and J. Helbert, *Spectrosc. Lett.* **47**, 446 (2014).